# Securing Digital Systems via Split-Chip Obfuscation

Joseph Sweeney, Samuel Pagliarini, Lawrence Pileggi
Dept. of Electrical and Computer Engineering
Carnegie Mellon University
Pittsburgh, PA 15213
Email: joesweeney@cmu.edu

*Abstract*—Security is an important facet of integrated circuit design for many applications. IP privacy and Trojan insertion are growing threats as circuit fabrication in advanced nodes almost inevitably relies on untrusted foundries. A proposed solution is Split-Chip Obfuscation that uses a combination trusted and untrusted IC fabrication scheme. By utilizing two CMOS processes, a system is endowed with the stronger security guaranties of a trusted legacy node while also leveraging the performance and density of an advanced untrusted node. Critical to the effectiveness of Split-Chip Obfuscation is finding an optimum partitioning of a system between the two ICs. In this paper, we develop a design flow for the Split-Chip Obfuscation scheme, defining the essential system metrics and creating a tool to rapidly assess the large design space. We demonstrate the concept of such a tool and show its application on an example SoC.

## I. Introduction

Due to increasing costs associated with IC (integrated circuit) technology scaling, the set of companies that manufacture third-party designs in state-of-the-art CMOS nodes has been reduced to a small number of internationally run foundries. In tandem, demonstrable security issues in the commercial space suggest that designs that require paramount security should not be fabricated in untrusted foundries [1]. Once a design is submitted for manufacture, the foundry is able to modify functionality by inserting Trojans or leak the design to third parties. Notable attempts to secure the manufacturing process include Logic-Locking, Programmable Logic, and Split-Manufacturing [2-4]. Each of these solutions increases security, but also has associated shortcomings.

Logic Locking obfuscates a circuit's functionality by adding keyed elements to a circuit. After manufacture, the designer enables the circuit by programing the key in an on-chip tamper-proof memory. The true functionality is not exposed to the foundry, hindering targeted Trojan insertion and IP leakage. Logic Locking however suffers from a metric problem wherein the security is quantified with heuristics [5]. Additionally, SAT-based attacks have been able to de-obfuscate locked designs. Programmable Logic has been used to withhold all or large portions of a design from the manufacturer. While this technique effectively obfuscates the design, it is associated with a large performance overhead, incompatible with some application requirements. Another technique, Split-Manufacturing splits a design into front- and back-end-of line parts. Each part is manufactured in a separate foundry and subsequently combined by a trusted entity. Thus, barring collusion, neither foundry has full access to the design. This is a potent solution but requires navigating complicated logistics between foundries.

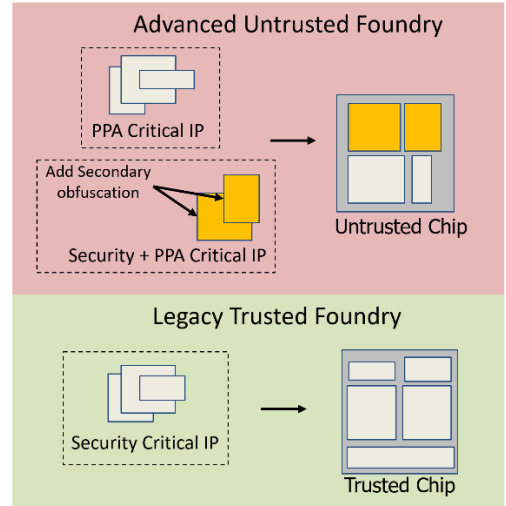

Fig. 1. Diagram of Split-Chip Obfuscation's dual foundry manufacturing. Performance and security critical IP are respectively place on the untrusted and trusted ICs. IP that is critical in both domains can be added to the untrusted IC with additional obfuscation.

An alternative approach to ensure a secure manufacturing process is Split-Chip Obfuscation, which utilizes a mixed fabrication scheme to provide combined security and performance that none of the aforementioned techniques can match. This obfuscation technique is targeted at systems with security constraints that justify the cost of a second IC. Fig. 1 depicts the Split-Chip Obfuscation flow. The original system is split between trusted and untrusted ICs. The trusted IC is manufactured in a controlled setting where strict oversight can verify that no are Trojans inserted and the design information is secure. The trusted IC will likely use a low cost, legacy node that is certifiably secure, thereby significantly impacting the performance, area, and power for that chip. Conversely, the untrusted IC is manufactured in a state-of-the-art node, but at an assumed adversarial foundry.

The system components are divided between the ICs depending on their respective security and performance constraints. Modules that are both security and performance critical can be placed on the untrusted chip with secondary obfuscation techniques. The trusted IC will most likely contain the system control and IP that is security critical, whereas the untrusted IC will serve as a computational farm, containing performance and power critical components. This dual IC paradigm affords the overall system with strong security

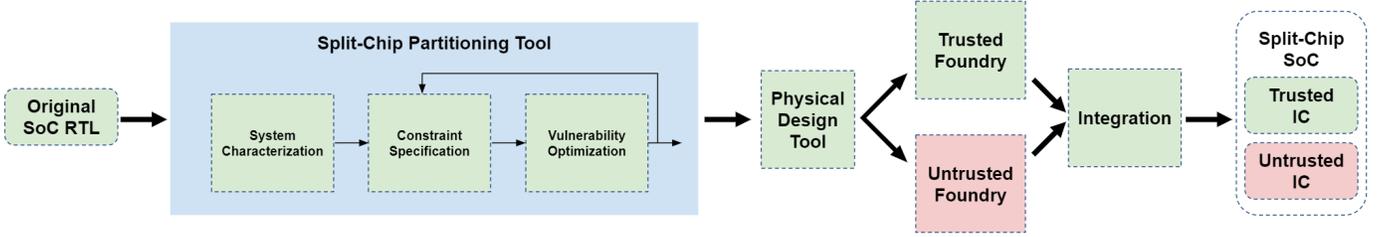

Fig. 2. Diagram of the Split-Chip Obfuscation flow with trusted entities in green and untrusted in red. The original SoC is split into two designs which are each produced using a standard IC flow. The designs are fabricated in separate foundries and integrated into a dual-chip, dual-trust SoC. The Split-Chip Partitioning Tool, the focus of this work, is highlighted in blue.

guarantees while maintaining performance.

Ideally, the two technologies nodes utilized in Spit-Chip allow for a system that achieves a trusted foundry's security while maintaining the performance of the untrusted foundry. However, in practice this is unlikely to be the case. The application of Split-Chip to an SoC requires careful planning on the part of the system engineer. The hard security guarantees do not come free. This integration process consists of partitioning the design between the two ICs such that the vulnerability to the untrusted foundry is minimized while still respecting the application dependent constraints. In using an older technology for a portion of the design, there will necessarily be a loss in both the system's power and area. Some trusted circuitry will be limited in its maximum frequency, which can reduce system throughput. Perhaps most impactful is the new communication bottleneck introduced between the two ICs. The system engineer must now consider all these additional effects while maximizing security within the bounds of the system constraints.

## II. SPLIT-CHIP PARTITIONING TOOL

As discussed in the previous section, the security benefits of Split-Chip are reliant on the intelligent integration of the trusted IC into a system. For simple systems, this process may be feasible by hand, but when extended to complex, heterogeneous systems, finding an optimal solution is intractable without automation. The number of possible partitions grows exponentially with the number of modules in the design and interactions between multiple interdependent metrics will not be immediately obvious to a designer. To assist a system designer in applying Split-Chip Obfuscation, we have developed a Split-Chip Partitioning tool (SCP) which builds a system model and quickly assesses the design space, finding the most secure yet feasible module partitioning. Additionally, the tool is able to add secondary obfuscation techniques to further secure the design. SCP gives a system engineer a flexible high-level interface to both integrate and evaluate Split-Chip Obfuscation in a system.

The SCP program flow consists of the following three steps:

### A. System Characterization:

SCP begins with system characterization, a onetime preprocessing step in which the parameters for the system model are determined. SCP targets the integration of Split-Chip Obfuscation into a System-on-Chip (SoC) design consisting of a set of functional modules connected with a Network on Chip (NoC), a paradigm pervasive in modern IC design. While the exact topology of the network may vary, we assume that internal module metrics are independent from the inter-module connections. As input to the characterization, the user specifies the RTL and connectivity of each module from the original SoC. The tool begins with an automated synthesis of each module in both the trusted and untrusted technology nodes. An iterative process is used to find the maximum frequency. Starting from zero, the clock period of each module is relaxed until the synthesis tool is able to converge on a solution that meets timing. The critical path slack is used to determine the next target clock period. This step produces netlists for each module in the trusted and untrusted configurations.

After synthesis, the untrusted module netlists are locked with secondary obfuscation, producing more possible module configurations. The locked netlists are again optimized by the synthesis tool to resize any modified gates. Currently, two forms of logic locking are applied: Key Logic, which uses keyed XOR gates to invert signals within the circuit [2], and FSM obfuscation, which adds decoy states to the circuit's FSM [4].

After this stage, each module has an associated netlist for four configurations: Trusted IC, Untrusted IC, Untrusted IC with Key Logic, and Untrusted IC with FSM Obfuscation. The resulting netlists are used to produce values for area, dynamic power, and static power at max frequency for each *module configuration*. These values then parameterize, the system model, a set of functions that produce metrics for a given *system configuration* which consists of a configuration assignment for each of the modules. While the time to synthesize each module in both technologies is high, this upfront overhead allows quick assessment and optimization in the later stages of the flow.

### B. Constraint Specification:

The system model consists of metrics for clock domain frequency, power, I/O bandwidth, I/O latency, and area. These metrics collectively capture the aspects of the system essential to adding Split-Chip Obfuscation. Each metric is constrained to a value determined by the user. Additionally, the user can specify hard placement constraints that fix a module to a given configuration. Constraint specification begins by assessing the initial space of possible configurations with the constraints set to the original SoC values. From this point, the constraints are iteratively relaxed, allowing modules to move from the untrusted IC to more secure configurations. This process continues until the user is satisfied with the resulting partitioning. In the remainder of the section, we discuss how each of these metrics is calculated.

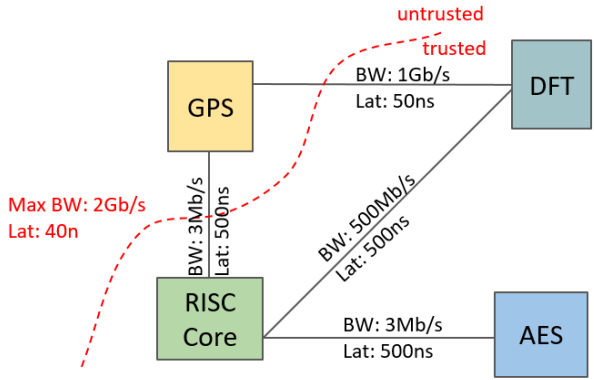

Fig. 3. Depiction of the graph representation used for constraining I/O bandwidth and latency. The boundary between ICs is shown with the red dashed line.

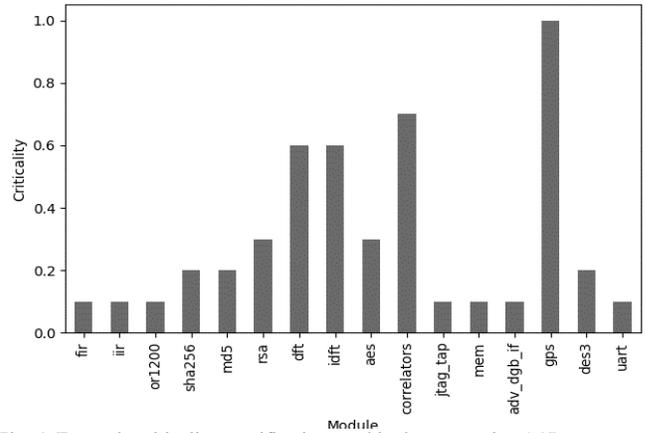

Fig. 4. Example criticality specification used in demonstrating SCP on an example SoC. The user uses criticality to rank the security importance of each module in the system.

*Clock Domain Frequency:* Typical SoCs are divided into multiple clock domains, linking sets of modules to the same frequency. This paradigm is replicated in SCP. For each clock domain, the user specifies a set of constituent modules and a minimum domain frequency. The clock domain frequency is calculated in two steps. First, the maximum frequency of each module in the domain is determined by the module's configuration. Next, the clock domain frequency is determined by the minimum of the module frequencies. Each module in the group is then run at this frequency.

*Power:* The next metric, power, builds upon the previously determined clock domain frequency. In the initial system characterization, the power at maximum frequency is found. However, in the final system, each module will not necessarily be run at this maximum. Thus, in order to estimate a module's power consumption at a given frequency, we scale the characterized dynamic power by the ratio between the module's maximum frequency and its clock domain frequency. While this estimate does not exactly correspond to what is obtained by resynthesizing the module with a new target frequency, it does give a close, and conservative estimate. This scaled dynamic power is added to the module's static power and contributions from each module are then totaled. The user is able to constrain the total system power as well the power for each individual IC.

*I/O Bandwidth:* Splitting an SoC between two ICs will inherently increase the total I/O bandwidth utilized by the system. Already in a single IC system, I/O bandwidth can be a critical bottleneck, limiting system throughput to the bandwidth provided by the specific I/O technology. In a Split-Chip system, the hard upper bound on I/O bandwidth must now include chip-to-chip communications that previously used an NoC, clearly making this a first order constraint. To model the system's I/O bandwidth, we create a graph representing the system, an example of which is shown in Fig. 3. Each module corresponds to a node with weighted edges representing the inter-module bandwidths. For each system configuration, we find the edges that are routed between the trusted and untrusted chips and sum their respective weights. This value is constrained to the maximum system I/O bandwidth.

*I/O Latency:* In addition to I/O bandwidth, the latency of off-chip connections must be considered. Any inter-module connection routed off-chip will now have to account for this delay. While some paths are expected to be indifferent to this latency change, others will have meet external, application dependent constraints. Similar to bandwidth, we use a graph representation to capture the latency requirements. The modules are represented as nodes and latencies between them are the edges. For any edge that is routed off-chip we increase the latency by the expected inter-chip delay. As shown in Fig. 3, these edge latencies are then constrained with user specified maximum latencies. We also allow the user to specify constraints that span multiple edges enabling more expressive security constraints.

*Area:* Area is an important metric depending on the system's application. We ensure that the total area of the configuration is less than the constraint specified. Similar to power, a module's area will change with its clock speed due to additional buffering requirements. This change is relatively small compared to the overall module area so we make the assumption that area remains constant with period.

### C. Vulnerability Optimization:

With the system model and constraints set, the user runs an optimization to find the most secure feasible system configuration. This process is naturally formulated as a discrete optimization problem, wherein each module is assigned one of four states: Trusted IC, Untrusted IC, Untrusted IC with Key Logic, or Untrusted IC with FSM Obfuscation. To weigh the relative securities of a given system configuration, we introduce a new metric, *vulnerability*. A higher vulnerability corresponds to a less secure system, thus the optimization will minimize vulnerability subject to the constraints. Vulnerability consists of the product of two user-specfied components: *exposure* and *criticality*.

Exposure assesses the risk associated with a module being in a particular configuration. The term represents the module's exposure to Trojan insertion or IP theft. We give the trusted chip a low exposure meaning the adversarial foundry does not have access to the modules. Meanwhile the untrusted chip is given an exposure of 1, following the assumption that the adversarial foundry has complete access. The secondary obfuscation techniques are given exposures close to 1 with their relative rankings determined by the designer's regard for each. The second vulnerability component, criticality, specifies the relative importance of the modules. Security critical modules are designated with a high criticality which will puch the optimization to prefer putting these modules in more secure configurations. In Fig. 4, we show the criticalities used for an example SoC.

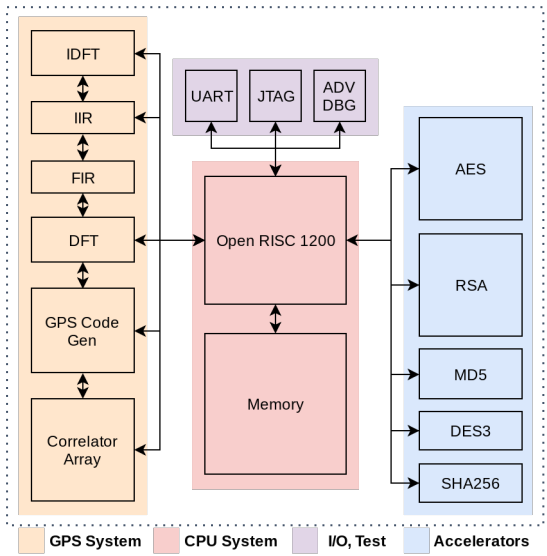

Fig. 5. Block diagram of example SoC with modules grouped by clock domain.

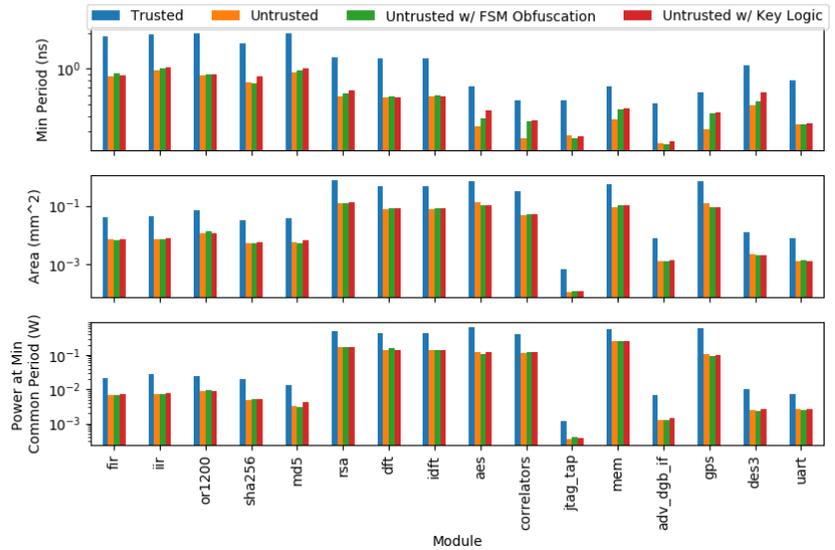

Fig. 6. System characterization results for each configuration for each module from the example SoC. Each module configuration is highlighted in a different color.

We stress the use of vulnerability as not a security guarantee, but simply a way to rank the various system configurations. Split-Chip's security still relies on the single hard guarantee of the foundry not having access to the trusted IC's netlist. Vulnerability is simply a metric that maximizes the use of this hard guarantee.

As previously mentioned, the space of possible system configurations scales exponentially with the number of modules in the design. Fortunately, the optimization problem solved by SCP has structure that makes it a good candidate for branch and bound optimization, allowing large systems to be solved quickly. Unlike a brute force approach, the branch and bound paradigm utilizes information from previous evaluations to eliminate portions of the search space. The branch and bound optimizer recursively considers sets of solutions, ruling out subsets based on a bounding function. The optimizer starts with the set of all solutions, in this case all possible module configurations. This set is divided into subsets by branching, recursively assigning modules to a given configuration. At each branch, the partial solution is evaluated. Because the metrics considered by our tool are additive, if the partition solution does not satisfy the constraints or produces a higher vulnerability than any known full solution, the subset of configurations in the branch is ruled out. This property allows the branch and bound paradigm to effectively prune the search space, greatly reducing the run time.

After the optimization is run, the results are displayed to the user as depicted in Fig. 7. Depending on the resulting vulnerability score, the user can accept the results or continue exploring the space by modifying the constraints. This process can be easily parallelized allowing quick parameter sweeps to assess the available tradeoffs.

### III. SoC Demonstration

To demonstrate the efficacy of SCP, we use it on an example SoC, the block diagram of which is displayed in Fig. 5. The system contains a total of 16 modules divided into four clock domains. The first domain contains a GPS system, which functions as the performance critical portion of the design. The six modules in this domain are interconnected with high bandwidth connections and expected to perform close to the nominal untrusted system. The next domain contains a set of cryptographic accelerators. An OpenRISC microprocessor and memory makes up the next domain. The microprocessor functions as the system's glue logic containing connections to all modules in the design and thus has relatively high performance constraints. Finally, the last domain contains I/O and test modules with relaxed security and performance constraints.

For the trusted and untrusted technologies, we respectively utilize 45nm and 16nm standard cell libraries. In Fig. 6, the results of the system characterization are displayed. The minimum period, area, and a normalized power value are shown for each module. As expected, across all modules there is a significant gap between the trusted and untrusted technology nodes. The two obfuscation techniques show similar average overheads, but the cheaper technique depends on the module. Even for a relatively simple system such as this, the wide range in results shows the complexity of partitioning a system by hand.

Fig. 7. displays four runs from a parameter sweep on our example system. The first two runs just use the trusted and untrusted configurations, whereas the third and fourth runs allow

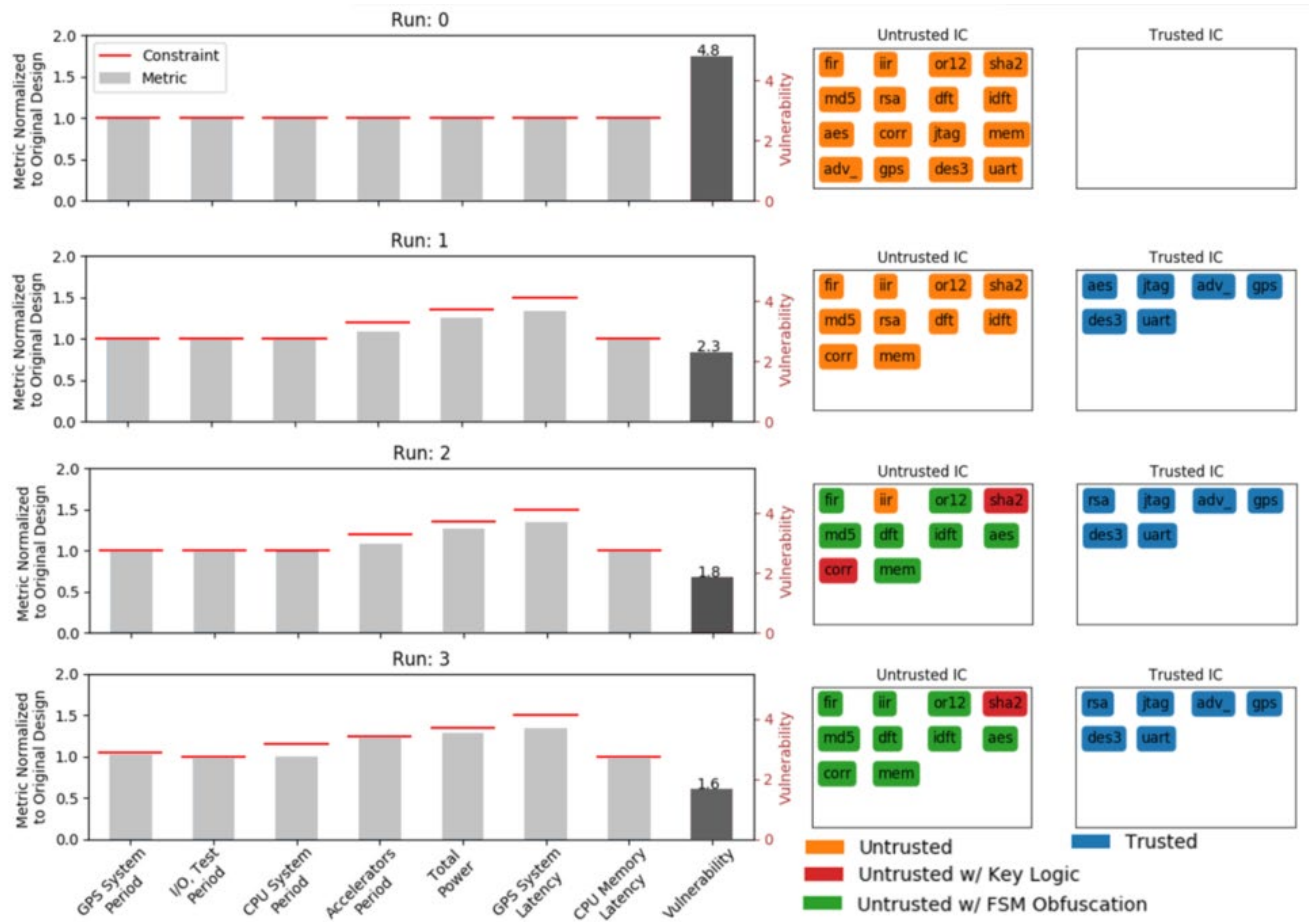

Fig. 7. Graphical interface of the Split-Chip Partitioning Tool, showing four runs from a constraint sweep. Each run shows a bar plot of the user-specified constraints and corresponding metrics for the optimized system. The vulnerability of each run is shown on the same plot with its axis in red. To the right, the partitioning is displayed with the modules colored based on their configuration.

secondary obfuscation to be added to the system. Displayed for each run are a subset of the system metrics, corresponding constraints, and partitioning diagram of the optimal solution.

In run 0, the system constraints have been held at the original SoC values. As a result, the tool is unable to find a secure solution. The only configuration that matches the constraints is all modules being placed on the untrusted IC. This run establishes a baseline vulnerability score that the user can compare to the subsequent results. In run 1, several constraints are relaxed namely the period of the accelerator clock domain, the total system power, and GPS system latency. This allows modules to be moved to the trusted IC, significantly reducing the system vulnerability. We see that the optimal solution moves several security critical modules to the trusted IC. In run 2, Logic-Locking and FSM obfuscation are enabled with the same relaxed constraints as the previous run. The result is a further reduction of the vulnerability from this secondary obfuscation. Finally in run 3, a small additional constraint relaxation on the CPU clock domain period, allows all modules in the system to have some form of obfuscation. While only a few iterations are shown, the user can continue this process to explore other potential tradeoffs.

## IV. CONCLUSION

Split-Chip Obfuscation shows great promise as a method of securely manufacturing low-volume ICs. However, to realize the potential benefits, careful attention must be payed to partitioning the system. To enable this process on complex designs we have developed SCP, which demonstrates how a powerful automation framework can explore the vast design space.

## ACKNOWLEDGMENT

This work was supported in part by the Defense Advanced Research Projects Agency under contract FA8750-17-1-0059 "Obfuscated Manufacturing for GPS (OMG)".